\documentstyle[pre,aps,twocolumn,amsfonts,eqsecnum]{revtex}
\begin{document}
\def\u{\bbox}
\def\Bbb{\relax}
\def\d{\displaystyle}
\def\mathcal#1{{\cal #1}}
\def\goldenmean{\gamma}
\def\phi{\varphi}
\def\epsilon{\varepsilon}
\def\goldenmean{\gamma}

\draft

\title{An Approximate KAM-Renormalization-Group Scheme for Hamiltonian Systems}

\author{C. Chandre$^1$, H. R. Jauslin$^1$, and G. Benfatto$^2$} 

\address{$^1$Laboratoire de Physique, CNRS, Universit\'e de Bourgogne,
BP 400, F-21011 Dijon, France}
\address{$^2$Dipartimento di Matematica, 
Universit\`a di Roma ``Tor Vergata'', 
Via della Ricerca Scientifica, I-00133 Roma, Italy}

\maketitle

\begin{abstract}
We construct an approximate renormalization scheme for Hamiltonian
systems with two degrees of freedom. This scheme is a combination of 
Kolmogorov-Arnold-Moser (KAM) theory 
and renormalization-group techniques. It makes the connection between the
approximate renormalization procedure derived by Escande and Doveil, and
a systematic expansion of the transformation. In particular, we show that 
the two main approximations, consisting in keeping only the quadratic
terms in the actions and the two main resonances, keep the essential
information on the threshold of the breakup of invariant tori.
\end{abstract}

\section{Introduction}

In 1981, Escande and Doveil~\cite{escandedoveil} set up an approximate
renormalization scheme for Hamiltonian systems with two degrees of
freedom, in order to study the breakup of invariant tori, and especially
to compute the threshold of stochasticity. Their scheme
was motivated by Chirikov's resonance overlap criterion~\cite{chirikov}, and by
Greene's results~\cite{greene} about the link between the existence  of a 
torus with the stability of neighboring periodic orbits. They established
the relevance of a sequence of these periodic orbits for the breakup of 
invariant tori, by setting an approximate transformation which focuses
successively on smaller scales, i.e. acting like a microscope in phase space.\\
Due to the complexity of the phase space of a non-integrable Hamiltonian, 
their method requires strong approximations to obtain explicit expressions.
Basically, two approximations were involved:\\
(1) a quadratic approximation in the actions: their transformation
produces terms that are higher than quadratic in the actions;
in order to remain in the same class of Hamiltonians, they neglect these
higher order terms,\\
(2) a two-resonance approximation: they only keep the two main resonances
at each iteration of the transformation.\\
The idea was to keep only the most relevant features of the mechanism of the 
breakup of a given torus.\\
In this article, we construct an approximate scheme using the same two
approximations. We establish the
connection between Escande's scheme~\cite{escandedoveil,escande} 
and the KAM-RG transformation derived in Refs.\ \cite{gallben,govin,chandre}.
The aim is to show that an exact renormalization transformation 
can be approximated by a 
simple transformation: It can be
useful to derive approximate explicit expressions of universal parameters, and
to see what are the most relevant terms responsible for the breakup of
invariant tori.
The results we obtain support the general idea that the irrelevant terms
of the renormalization
transformation can be eliminated with little loss of accuracy in the
parameters associated to the breakup of
invariant tori.\\
The transformation $\mathcal{R}$ we define has two main parts: a KAM 
transformation which
is a canonical change of coordinates that reduces the size of the perturbation
from $\epsilon$ to $\epsilon^2$,
and a renormalization transformation which is a combination of a shift of
the resonances and a rescaling of momentum and energy.\\
It acts on the following class of Hamiltonians with two degrees of freedom, 
quadratic in the action variables $\u{A}=(A_1,A_2)$, and described by three
even scalar functions of the angles $\u{\phi}=(\phi_1,\phi_2)$:
\begin{eqnarray}
  H(\u{A},\u{\phi})=&&\frac{1}{2}\left(1+m(\u{\phi})\right)
        (\u{\Omega}\cdot\u{A})^2 \nonumber \\
	&& +\left[\u{\omega}_0+g(\u{\phi})\u{\Omega}
	\right]\cdot\u{A}+f(\u{\phi}),\label{hamiltonian}
\end{eqnarray}
where $m$, $g$, and $f$ are of zero average.
The vector $\u{\omega}_0$ is the frequency vector of the considered torus and
$\u{\Omega}=(1,\alpha)$ is some other constant vector not parallel to
$\u{\omega}_0$. 
The perturbation $(m,g,f)$ is of order $O(\varepsilon)$. \\
The renormalization-group approach is based on the following general picture:
The idea is to construct the transformation $\mathcal{R}$ 
as a generalized canonical 
change of coordinates acting on some space of Hamiltonians such that the 
iteration of $\mathcal{R}$ converges to a fixed point. If the 
perturbation is smaller than critical,
$\mathcal{R}$ should converge to a Hamiltonian of type~(\ref{hamiltonian})
with $(m,g,f)=0$, which is integrable, and the equations
of motion show that the torus with frequency vector
$\u{\omega}_0$ is located at $\u{A}=0$.
All Hamiltonians attracted by this trivial fixed point have an
invariant torus
of that frequency (this can be considered as an
alternative version of the KAM theorem~\cite{koch}).
If the perturbation is larger than critical, the system does not have a KAM 
torus of the considered frequency and the iteration of $\mathcal{R}$ diverges.
The domain of 
convergence to the trivial fixed point and the domain of divergence 
are separated by a {\it critical surface}
invariant under the action of $\mathcal{R}$. The main hypothesis of the 
renormalization-group approach is that there should be another nontrivial 
fixed point (or more generally, a fixed set) on this critical surface, 
that is attractive for Hamiltonians 
on that surface. From its existence, one can expect to deduce universal 
properties in the mechanism of the breakup of invariant tori.\\
Many aspects of this general picture are still at the stage of conjecture,
supported by some results in the perturbative regime \cite{koch}, by
numerical works \cite{govin,chandre,cgjk,abad}, and by analogies with
the related problem for area-preserving maps \cite{mackay,mackayL}. 
In particular, the relation between the properties of the nontrivial
renormalization fixed point and the geometric properties of the invariant 
torus at the instability threshold are not well established. The
coincidence of the critical coupling of one-parameter families at 
which a torus breaks up, with the boundary of attraction of the trivial
fixed point  is supported by numerical studies \cite{govin,chandre,cgjk}.\\

In Sec.\ \ref{sec:kam}, we describe the KAM part of the transformation, 
and we make explicit
the two approximations involved in this scheme: the quadratic approximation
and the two-resonance approximation.
In Sec.\ \ref{sec:ren}, we present the
renormalization transformation which is a combination
of the KAM part, a shift of the resonances, 
and rescalings of actions and energy. 
In Sec.\ \ref{sec:res}, we give
our numerical results, and in particular, we show that the approximate scheme 
contains the essential features of the exact one. 

\section{KAM transformation}
\label{sec:kam}

We perform a canonical transformation $\mathcal{U}_F:(\u{\phi},\u{A})
\mapsto (\u{\phi}',\u{A}') $  
defined by a generating function 
$F(\u{A}',\u{\phi})$~\cite{goldstein,gallavottiL} 
characterized by a
scalar function $X$ of the action and angle variables, of the form
\begin{equation}
F(\u{A}',\u{\phi})=\u{A}'\cdot\u{\phi}+ X(\u{A}',\u{\phi}),
\end{equation}
leading to
\begin{eqnarray}
&&  \label{eqn:tcan1}
    \u{A}=\d \frac{\partial F}{\partial \u{\phi}}
	 =\u{A}'+\frac{\partial X}{\partial \u{\phi}},\\
&& \label{eqn:tcan2}
    \u{\phi}'=\d \frac{\partial F}{\partial \u{A}'}
             =\u{\phi}+\frac{\partial X}{\partial \u{A}'}.
\end{eqnarray}
The function $X$ is constructed such that in
$\mathcal{H}\circ\mathcal{U}_F$ the perturbation terms of first order 
in $\epsilon$ are equal to zero.
Inserting Eq.\ (\ref{eqn:tcan1}) into Hamiltonian (\ref{hamiltonian}), one
obtains the expression of the Hamiltonian
in the mixed representation of new action variables and old angle variables:
\begin{eqnarray}
\tilde{H}({\u A}',{\u \varphi})=&&(\u{\Omega}\cdot\u{A}')^2/2
+ {\u \omega}_0 \cdot {\u A}'\nonumber \\
&& + \u{\omega}(\u{A}')\cdot\frac{\partial X}{\partial \u{\phi}}+ 
         h(\u{A}',\u{\phi}) +O(\varepsilon^2),
\end{eqnarray}
where
\begin{eqnarray}
    && \u{\omega}(\u{A}')=
	 \u{\omega}_0+(\u{\Omega}\cdot\u{A}') \u{\Omega},\\
    && h(\u{A}',\u{\phi})=\frac{1}{2}m(\u{\phi})(\u{\Omega}\cdot\u{A}')^2+
         g(\u{\phi})\u{\Omega}\cdot\u{A}'+f(\u{\phi}).
\end{eqnarray}
The equation that determines $X$ is thus:
\begin{equation}
\label{eqn:eqX}
 \u{\omega}(\u{A}')\cdot\frac{\partial X}{\partial \u{\phi}}+ 
         h(\u{A}',\u{\phi})=0.
\end{equation}
We recall that the functions $m$, $g$, and $f$ are of order 
$O(\epsilon)$; as a consequence, $X$ is also of order 
$O(\epsilon)$.
Equation (\ref{eqn:eqX}) has the solution 
\begin{equation}
\label{eqn:X}
  X(\u{A}',\u{\phi})=\sum_{\nu \in {\Bbb Z}^2} 
  X_{\nu}(\u{A}') \sin(\u{\nu}\cdot\u{\phi}),
\end{equation}
where, if we write $h(\u{A},\u{\phi})=\sum_{\nu} h_{\nu}(\u{A}) 
\cos(\u{\nu}\cdot\u{\phi})$,
\begin{equation}
\label{eqn:Xnu}
X_{\nu}(\u{A}')=-\d \frac{h_{\nu}(\u{A}')}{\u{\omega}(\u{A}')\cdot\u{\nu}}.
\end{equation}

The denominator of $X_{\nu}$ depends on the actions: thus, by power
expansion, it generates terms 
that are 
higher than quadratic in the actions.
In order to remain in the same space of Hamiltonians~(\ref{hamiltonian}),
we expand the Hamiltonian to
the second order in the actions and neglect the order $O(A^3)$.
The justification for such an approximation is that we are interested 
in the torus with frequency vector $\u{\omega}_0$ 
which is located at $\u{A}=0$ for the trivial fixed point.\\
We consider Hamiltonians~(\ref{hamiltonian}) with only two Fourier modes
which are the two main resonances defined as follows: 
For a frequency vector $\u{\omega}_0$, the resonances are given by 
the vectors $\u{\nu}_n=(p_n,q_n)$ which are the sequence of the best
rational approximations. They are characterized precisely
by the following property:
$|\u{\omega}_0\cdot\u{\nu}_n| < |\u{\omega}_0\cdot\u{\nu}|$, for any
$\u{\nu}\equiv (p,q)\not=\u{\nu}_n$ such that $|q|<q_{n+1}$.
For the frequency vector
$\u{\omega}_0=(1/\gamma,-1)$ with $\gamma=(1+\sqrt{5})/2$,
we define the two {\it main resonances} as $\u{\nu}_1=(1,0)$ and
$\u{\nu}_2=(1,1)$. Well known
properties of continued fractions (see, for example, \cite{davenport}) imply
that the vectors $\u{\nu}_n$ satisfy the recursion relation
$\u{\nu}_{n+1}=N \u{\nu}_n$, where
$N =\d\left(\begin{array}{cc} 1 & 1 \\ 1 & 0 \end{array}\right)$.\\
The scalar functions $m$, $g$ and $f$  will have expressions of the form
\begin{equation}
f(\u{\phi})=f_{\nu_1}\cos(\u{\nu}_1\cdot\u{\phi})
            +f_{\nu_2}\cos(\u{\nu}_2\cdot\u{\phi}).
\end{equation}
The KAM part of the transformation $\mathcal{R}$ will generate
a large set of Fourier modes from these two main resonances. 
The purpose of the renormalization is to relate these two main resonances 
representing the 
main scale, to the next pair of resonances 
(called the {\it daughter resonances})
representing the next smaller scale. 
For the frequency vector
$\u{\omega}_0=(1/\gamma,-1)$, the daughter resonances are $\u{\nu}_3=(2,1)$ 
and $\u{\nu}_4=(3,2)$.
As the canonical transformation (\ref{eqn:tcan1})-(\ref{eqn:tcan2}) is linear
in $\cos(\u{\nu}\cdot\u{\phi})$ and $\sin(\u{\nu}\cdot\u{\phi})$,
$\u{\nu}_n=\u{\nu}_{n-1}+\u{\nu}_{n-2}$ and
the three scalar functions $(m,g,f)$ are of
order $O(\varepsilon)$, the lowest order to which the resonances
$\u{\nu}_3$ and $\u{\nu}_4$ are produced, is respectively
$O(\varepsilon^2)$ and $O(\varepsilon^3)$. Thus we
neglect the order $O(\varepsilon^4)$ of the KAM transformation.\\
The next step is to express the Hamiltonian in the new angle variables using
Eq.\ (\ref{eqn:tcan2}). We notice that this equation has to be inverted. As we
need the Hamiltonian expressed in the new coordinates to order
$O(\epsilon^3)$, we have to invert Eq.\ (\ref{eqn:tcan2}) up to  
order 
$O(\epsilon)$, since the functions to be expressed in the new angles
are already of order $O(\epsilon^2)$. Moreover, these new angle 
variables depend on the actions, which we develop to order 
$O(A^2)$. The development can thus be summarized by expressing
$\cos[\u{\nu}.\u{\phi}(\u{\phi}')]$ as a 
function of $\cos(\u{\nu}.\u{\phi}')$ and 
$\sin(\u{\nu}.\u{\phi}')$, neglecting the orders 
$O(\epsilon^2,A^3)$.\\
The final step is a translation in the ${\u A'}$ variables such that the linear term
is again of the form $\u{\omega}_0\cdot \u{A}'$.\\
The Hamiltonian expressed in the new 
variables becomes:
\begin{eqnarray}
  H'(\u{A}',\u{\phi}')=&& \frac{1}{2}\left(\Lambda+m'(\u{\phi}')\right)
        (\u{\Omega}\cdot\u{A}')^2 \nonumber \\
	&&+\left(\u{\omega}_0+g'(\u{\phi}')
	\u{\Omega}\right)\cdot\u{A}'+f'(\u{\phi}'),\label{hamimage}
\end{eqnarray}
where $m'$, $g'$ and $f'$ are given as functions
of $m$, $g$ and $f$. The constant $\Lambda$ results of the fact that the 
mean value of $m'$ is required to be equal to zero. We notice that the KAM 
transformation does not change $\u{\Omega}=(1,\alpha)$.\\
The two-resonance approximation consists in
retaining only the two daughter resonances and neglect all the other
Fourier modes.
Therefore the approximate KAM transformation
$\tilde\mathcal{U}_F$ is a map acting
on a low-dimensional space of Fourier coefficients:
\begin{eqnarray}
&& \tilde\mathcal{U}_F\left(1;m_{\nu_1},g_{\nu_1},f_{\nu_1};m_{\nu_2},
  g_{\nu_2},f_{\nu_2}\right) \nonumber \\
&& \qquad  =\left(\Lambda;m'_{\nu_3},g'_{\nu_3},f'_{\nu_3};m'_{\nu_4},
  g'_{\nu_4},f'_{\nu_4}\right).
\end{eqnarray}
The explicit expression of this map is given in the Appendix.

\section{Renormalization transformation}
\label{sec:ren}

We construct the approximate transformation by combining two parts: 
a KAM transformation $(m,g,f,\alpha)\mapsto(m',g',f',\alpha)$ as defined above, 
and a renormalization (RG) consisting of a shift of the resonances and a rescaling
of the actions and of time $(m',g',f',\alpha)\mapsto(m'',g'',f'',\alpha')$.
The renormalization scheme described in this section
is for a torus of frequency vector $\u{\omega}_0=(1/\goldenmean,-1)$ where
$\goldenmean=(1+\sqrt{5})/2$. It is straightforward to adapt it to quadratic
irrationals.\\
The approximate KAM-RG transformation is composed of four steps:\\
1) a KAM transformation described in the previous section,
which is a change of coordinates that eliminates
terms of order $O(\epsilon)$, where $\epsilon$ is the size of the
perturbation; this transformation produces terms of order 
$O(\epsilon^2)$, terms that are higher than quadratic in the 
actions, and a large set of Fourier modes.
We neglect terms of order 
$O(A^3,\epsilon^4)$, and also, all the Fourier modes except
the two daughter resonances $\u{\nu}_3$ and $\u{\nu}_4$. 
We notice that this transformation 
does not change $\u{\Omega}$.\\
2) a shift of the resonances: a canonical change of coordinates that maps 
the pair of daughter resonances $(\u{\nu}_3,\u{\nu}_4)$
into the two main resonances $(\u{\nu}_1,\u{\nu}_2)$.\\
3) a rescaling of energy (or equivalently of time).\\
4) a rescaling of the action variables (which is
	      a generalized canonical transformation).\\
The aim of this transformation is to treat one scale at the time. 
The steps 2), 3) and 4) are implemented as follows:
The two main resonances $(1,0)$ and $(1,1)$ are replaced by the 
next pair of daughter resonances $(2,1)$ and $(3,2)$, i.e. we require
that $\cos[(2,1)\cdot\u{\phi}']=\cos[(1,0)\cdot\u{\phi}'']$ and 
$\cos[(3,2)\cdot\u{\phi}']=\cos[(1,1)\cdot\u{\phi}'']$.
This change is done via a canonical transformation 
$(\u{A}',\u{\varphi}')\mapsto (N^{-2}\u{A}',N^2\u{\varphi}')$ with 
$$N^2=\d\left(\begin{array}{cc} 2 & 1 \\ 1 & 1 \end{array}\right).$$
This linear transformation multiplies ${\u \omega}_0$ by $\goldenmean^{-2}$
(since $\u{\omega}_0$ is an eigenvector of $N$);
therefore we rescale the energy by a factor $\goldenmean^2$ in order to keep
the frequency fixed at ${\u \omega}_0$. A consequence of the shift of
the resonances is that $\u{\Omega}$ is changed into $\u{\Omega}'=(1,\alpha')$,
where $\alpha'=(\alpha+1)/(\alpha+2)$.\\
Then we perform a rescaling of the action variables: we change the  
Hamiltonian $H'$ into $$\hat{H'}({\u A}',{\u \varphi}')=
\lambda H'\left(\d\frac{{\u A}'}{\lambda},{\u \varphi}'\right)$$ with $\lambda$
such that Hamiltonian (\ref{hamimage}) becomes of the form (\ref{hamiltonian}).
Since the rescaling of energy and the shift $N^2$ transform the 
quadratic term of the Hamiltonian into 
$\goldenmean^2(2+\alpha)^2[\Lambda+m'(\u{\phi}')](\u{\Omega}'\cdot 
\u{A}')^2/2$, this condition leads to
$\lambda=\goldenmean^2(2+\alpha)^2\Lambda$.
This condition has the following geometric interpretation in terms of 
self-similarity of the resonances close to the invariant torus:
the rescaling magnifies the size of the daughter resonances,
and places them approximately at the location of the original 
main resonances~\cite{chandre}.\\
In summary, the renormalization rescales $m$, $g$, $f$ and 
$\u{\Omega}=(1,\alpha)$ into
\begin{eqnarray}
     && m''(\u{\phi})=\d \frac{m'\left(N^{-2}\u{\phi}\right)}
                     {\Lambda},\\
     && g''(\u{\phi})=\goldenmean^2(2+\alpha) g'\left(N^{-2}\u{\phi}
                      \right),\\
     && f''(\u{\phi})=\goldenmean^4(2+\alpha)^2 \Lambda
                     f'\left(N^{-2}\u{\phi}\right),\\
     && \alpha'=\frac{1+\alpha}{2+\alpha}.\label{eqn:alpha}
\end{eqnarray}
The iteration of the transformation (\ref{eqn:alpha}) converges to 
$\alpha_*=\goldenmean^{-1}$. It means that ${\u \Omega}$ converges under 
successive iterations  to ${\u \Omega}_*=(1,1/\goldenmean)$, which
 is orthogonal to ${\u \omega}_0$ and is the unstable 
eigenvector of $N^2$ with the largest eigenvalue $\goldenmean^2$.

\section{Determination of the critical coupling; an
          approximate nontrivial fixed point}
\label{sec:res}
	  
We start with the same initial Hamiltonian as in
Refs.\ \cite{escande,govin,chandre} 
\begin{equation}
\label{hamiltonieninit}
H({\u A},{\u \varphi})=\frac{1}{2}({\u \Omega}\cdot{\u A})^{2}
+{\u \omega}_{0}\cdot{\u A}+\varepsilon f({\u \varphi}) \ ,
\end{equation}
where ${\u \Omega}=(1,0)$, ${\u \omega}_0
=(1/\goldenmean,-1)$,
$\goldenmean=(1+\sqrt{5})/2$, and a perturbation
\begin{equation}
f({\u \varphi})=\cos({\u \nu}_1\cdot{\u \varphi})
+ \cos({\u \nu}_2\cdot{\u \varphi}), 
\end{equation}
where ${\u \nu}_1=(1,0)$ and ${\u \nu}_2=(1,1)$.\\
We take successively larger coupling $\epsilon$, and determine whether
the approximate KAM-RG iteration converges to a Hamiltonian with $(m,g,f)=0$, or
whether it diverges $(m,g,f)\rightarrow \infty$. By a bisection procedure, we
determine the critical coupling $\epsilon_c=0.02885$. As a comparison, 
Escande's scheme~\cite{escande} gives $\epsilon_c=0.02908$.
The KAM-RG transformation~\cite{govin,chandre,cgjk} yields $\epsilon_c=0.02759$
(Greene's criterion gives also $\varepsilon_c=0.02759$).
The two-resonance scheme gives the result within $5\%$. Thus, the
approximate scheme gives a fairly good description of
the critical surface of the breakup of the torus.\\
The renormalization operator has two fixed points: a trivial fixed point which 
corresponds to the Hamiltonian 
$H(\u{A},\u{\phi})
=(\u{\Omega}_*\cdot\u{A})^2/2+\u{\omega}_0\cdot\u{A}$,
and a nontrivial fixed point which lies on the boundary of the basin of 
attraction
of the trivial fixed point.
The critical surface which is the stable manifold of the nontrivial 
fixed point is of codimension
1. The relevant critical exponent is $\delta=2.7135$. This value is close
to the one obtained by Escande \textit{et al.}~\cite{escandeetal}
$\delta=2.7480$, or by KAM-RG schemes \cite{cgjk,abad} 
$\delta=2.6502$,
and by MacKay for area-preserving 
maps~\cite{mackay} $\delta=2.6502$.\\
The fact that $\delta$ is close to $\gamma^2$ can be understood by the following
heuristic arguments: To the first main resonance $\u{\nu}_1$ corresponds a 
Fourier component
$M \exp [i(1,0)\cdot\u{\varphi}]$, and to the second one
$P \exp [i(1,1)\cdot\u{\varphi}]$. 
The first daughter resonance $\u{\nu}_3$ is represented by 
$M'\exp [i(2,1)\cdot\u{\varphi}]$. 
The transformation is a polynomial change of coordinates that generates a set
of Fourier modes from the two main resonances. The way to generate the first
daughter resonance at the lowest order
is to combine one resonance $\u{\nu}_1$ and one resonance
$\u{\nu}_2$. For the second daughter resonance, the change of coordinates
must combine one resonance $\u{\nu}_1$ and two resonances $\u{\nu}_2$.
We rescale the phase space in such a way that the daughter resonances become
of the same size as the main resonances. 
These arguments give the following renormalization 
relations:
\begin{eqnarray}
&& M'=k_1 M P,\\
&& P'=k_2 M P^2,
\end{eqnarray}
where $k_1$ and $k_2$ are two constants that depend on how the transformation
is performed. \\
An analysis of this scheme shows that it has two fixed points:
a trivial one $M=0,\,P=0$ and a nontrivial one $M_*=k_1 k_2^{-1},\,
P_*=k_1^{-1}$. The nontrivial fixed point has a stable manifold
of codimension 1 characterized by a relevant critical exponent; the only
eigenvalue greater than one of the linearized map at the nontrivial fixed 
point is $\delta=\goldenmean^2$. This relevant
critical exponent does not depend on $k_1$ and $k_2$. This is in
agreement with the general ideas of the renormalization group.
Also, it shows that the relevant critical 
exponent of the approximate and exact KAM-RG transformation
should be expected to be close to $\goldenmean^2$.\\ 
For the scaling factor at the nontrivial fixed point, we obtain numerically
$\lambda_*=19.1248$. This value can be compared with $\lambda_*=18.8282$ 
obtained in Refs.\ 
\cite{kadanoff,shenkerkadanoff,mackay,cgjk,abad}. Table 1 lists some of 
the universal 
parameters associated with the breakup of golden tori.\\
The approximate renormalization transformation has another fixed set
which is a cycle of period three as it had also been encountered in 
area-preserving maps~\cite{mackayL,greenemao} and in the KAM-RG 
transformation~\cite{chandre}. This cycle is simply related to the nontrivial
fixed point by symmetries. In particular, it belongs to the same universality
class as the fixed point.

\section*{acknowledgments}

We thank A. Celletti for providing the program 
to determine the critical coupling by Greene's criterion. We acknowledge
useful discussions with G. Gallavotti, R. S. MacKay, and H. Koch.
Support from EC contract No.\ ERBCHRXCT94-0460 for the project
``Stability and universality in classical mechanics''
and from  the Conseil R\'egional de Bourgogne is acknowledged.

\newpage
\section*{Appendix: two-resonance scheme formulae}

We denote $c_{\nu}=\cos(\u{\nu}\cdot\u{\phi})$, 
$c_{\nu}'=\cos(\u{\nu}\cdot\u{\phi}')$,
and $s_{\nu}'=\sin(\u{\nu}\cdot\u{\phi}')$.
The expression of the Hamiltonian in the mixed representation of new actions
and old angles is 
\begin{eqnarray}
\tilde{H}(\u{A}',\u{\phi})&=&\frac{1}{2}(\u{\Omega}\cdot\u{A}')^2+\u{\omega}_0
\cdot\u{A}' \nonumber \\
&& +\sum_{\nu_1,\nu_2}P_{\nu_1\nu_2}(\u{A}') c_{\nu_1}c_{\nu_2} \nonumber \\ 
&& +\sum_{\nu_1,\nu_2,\nu_3} Q_{\nu_1\nu_2\nu_3}(\u{A}') 
c_{\nu_1}c_{\nu_2}c_{\nu_3},
\end{eqnarray}
where
\begin{eqnarray}
P_{\nu_1\nu_2}(\u{A})=\u{\Omega}\cdot\u{\nu}_1 X_{\nu_1}(\u{A})&&\left[
\u{\Omega}\cdot\u{\nu}_2 X_{\nu_2}(\u{A})/2\right.\nonumber \\
&&\left. +g_{\nu_2}+m_{\nu_2}\u{\Omega}\cdot
\u{A} \right],
\end{eqnarray}
\begin{equation}
 Q_{\nu_1\nu_2\nu_3}(\u{A})=\u{\Omega}\cdot\u{\nu}_1 \u{\Omega}\cdot\u{\nu}_2
X_{\nu_1}(\u{A}) X_{\nu_2}(\u{A}) m_{\nu_3}/2.
\end{equation}
We recall that $X$ is of order $O(\epsilon)$, and 
is given by Eq.\ (\ref{eqn:Xnu}); as a consequence,
$P$ is of order $O(\epsilon^2)$, and $Q$ is of order 
$O(\epsilon^3)$.\\
Next, the expression of this Hamiltonian in the new angles requires the 
inversion
of Eq.\ (\ref{eqn:tcan2}) to the order $O(\epsilon)$.
The expression of $c_{\nu}$ as a function
of $s_{\nu}'$ and $c_{\nu}'$ is 
\begin{equation}
c_{\nu_1}=c_{\nu_1}'+\sum_{\nu_2}R_{\nu_1\nu_2}(\u{A}') s_{\nu_1}'s_{\nu_2}'
+O(\epsilon^2),
\end{equation}
where $R_{\nu_1\nu_2}(\u{A})=\u{\nu}_1\cdot\d \frac{\partial X_{\nu_2}}{\partial
\u{A}}$.\\
The Hamiltonian (\ref{hamiltonian}) expressed in the new variables becomes 
\begin{eqnarray}
\label{hamil3}
H'(\u{A}',\u{\phi}')&=&\frac{1}{2}(\u{\Omega}\cdot\u{A}')^2+\u{\omega}_0
\cdot\u{A}' \nonumber \\
&& +\sum_{\nu_1,\nu_2} P_{\nu_1\nu_2}c_{\nu_1}'c_{\nu_2}'\nonumber \\
&& + 
 \sum_{\nu_1,\nu_2,\nu_3} Q_{\nu_1\nu_2\nu_3}
c_{\nu_1}'c_{\nu_2}'c_{\nu_3}' \nonumber \\
&& +\sum_{\nu_1,\nu_2,\nu_3}(P_{\nu_1\nu_2}+P_{\nu_2\nu_1})R_{\nu_2\nu_3}
c_{\nu_1}'s_{\nu_2}'s_{\nu_3}'\nonumber \\
&& + O(\epsilon^4).
\end{eqnarray}
The next approximation is the Taylor expansion of $H'(\u{A}',\u{\phi}')$ to the
second order in the actions e.g.
\begin{equation}
P_{\nu_1\nu_2}(\u{A})=P_{\nu_1\nu_2}^{(0)}+P_{\nu_1\nu_2}^{(1)}\u{\Omega}\cdot
\u{A}+P_{\nu_1\nu_2}^{(2)}(\u{\Omega}\cdot\u{A})^2 +O(A^3).
\end{equation}
In the next step, we neglect all the resonances different from the daughter
resonances $(2,1)$ and $(3,2)$ using the following relations
\begin{equation}
    c_{\nu_1}' c_{\nu_2}'=\frac{1}{2}\left(c_{\nu_1+\nu_2}'
         +c_{\nu_1-\nu_2}'\right),
\end{equation}
\begin{eqnarray}
c_{\nu_1}'s_{\nu_2}'s_{\nu_3}'=\frac{1}{4}
 &&\left(c_{\nu_1+\nu_2-\nu_3}'+c_{\nu_1-\nu_2+\nu_3}' \right. \nonumber \\
 && \left. -c_{\nu_1+\nu_2+\nu_3}'-c_{\nu_1-\nu_2-\nu_3}'\right),
\end{eqnarray}
\begin{eqnarray}
c_{\nu_1}'c_{\nu_2}'c_{\nu_3}'=\frac{1}{4}&&\left(c_{\nu_1+\nu_2-\nu_3}'
        +c_{\nu_1-\nu_2+\nu_3}'\right. \nonumber \\
&& \left. +c_{\nu_1+\nu_2+\nu_3}'+c_{\nu_1-\nu_2-\nu_3}'
	\right).
\end{eqnarray}
In the following formulas, the subscripts 1, 2, 3 and 4 denote respectively
the resonances $(1,0)$, $(1,1)$, $(2,1)$, and $(3,2)$. The Hamiltonian 
(\ref{hamil3})
becomes 
\begin{eqnarray}
\label{hamil4}
H'(\u{A}',\u{\phi}')&=&\frac{1}{2}(\u{\Omega}\cdot\u{A}')^2+\u{\omega}_0
\cdot\u{A}' +\frac{1}{2}(P_{11} +P_{22}) \nonumber \\
&& + \frac{1}{2}(P_{12}+P_{21}) c_{\nu_3}'\nonumber \\
&& +\frac{1}{4}(Q_{221}+Q_{212}+
Q_{122}-S_{122})c_{\nu_4}',
\end{eqnarray}
where $S_{122}=2P_{22}R_{21}+(P_{21}+P_{12})(R_{12}+R_{22}).$ Equation 
(\ref{hamil4})
gives the expression of the Fourier coefficients of the daughter resonances
\begin{eqnarray}
&& \Lambda=1+P_{11}^{(2)}+P_{22}^{(2)},\\
&& m_{\nu_3}'=P_{12}^{(2)}+P_{21}^{(2)},\\
&& m_{\nu_4}'=\frac{1}{2}\left(Q_{221}^{(2)}+Q_{212}^{(2)}+Q_{122}^{(2)}
-S_{122}^{(2)}\right),\\
&& g_{\nu_3}'=\frac{1}{2}\left(P_{12}^{(1)}+P_{21}^{(1)}\right),\\
&& g_{\nu_4}'=\frac{1}{4}\left(Q_{221}^{(1)}+Q_{212}^{(1)}+Q_{122}^{(1)}
-S_{122}^{(1)}\right),\\
&& f_{\nu_3}'=\frac{1}{2}\left(P_{12}^{(0)}+P_{21}^{(0)}\right),\\
&& f_{\nu_4}'=\frac{1}{4}\left(Q_{221}^{(0)}+Q_{212}^{(0)}+Q_{122}^{(0)}
-S_{122}^{(0)}\right).
\end{eqnarray}

\onecolumn
\begin{table}
\caption{Universal parameters associated with the breakup of the
golden mean torus.}
\begin{tabular}{c c c c}
Name & Value \tablenote{given in Refs.\ \cite{mackayL,escande}}& 
Approximate scheme & Simple scheme \tablenote{derived in Refs.\ 
\cite{escande,escandeetal}}\\
\hline
Unstable eigenvalue & 2.6502 & 2.7135 & $\gamma^2\approx 2.6180$ \\
Largest stable eigenvalue & 0.3731 & 0.4087 & $\gamma^{-2}\approx 0.3820$ \\
Area multiplier $\lambda_*$ & 18.8282 & 19.1248 & $\gamma^6\approx 17.9443$ \\
Inverse mass multiplier $\Lambda_*$ & 1.0493 & 1.0658 & 1 \\
Time multiplier & $\gamma^{-2}$ & $\gamma^{-2}$ & $\gamma^{-2}$ \\
\end{tabular}
\end{table}


\end{document}